\documentclass[twocolumn,aps,prx,floatfix,superscriptaddress]{revtex4-1}
\usepackage{amsmath,amssymb}
\usepackage{graphicx,float}
\usepackage{times,txfonts}
\usepackage{color}
\usepackage{multirow}
\usepackage{bbold}
\usepackage{color}
\usepackage{soul}

\usepackage{hyperref}
\usepackage{times,txfonts}
\usepackage{float}
\usepackage{natbib}
\usepackage{nicefrac}
\usepackage{blindtext}
\usepackage[export]{adjustbox}
\usepackage{mathrsfs}
\usepackage{textcomp}
\usepackage{blkarray}
\usepackage{multirow}

\newcommand{\bra}[1]{\left\langle #1 \right|}
\newcommand{\ket}[1]{\left| #1 \right\rangle}

\renewcommand{\epsilon}{\varepsilon}

\def\VR{\kern-\arraycolsep\strut\vrule &\kern-\arraycolsep}
\def\vr{\kern-\arraycolsep & \kern-\arraycolsep}
\usepackage{sistyle}
\usepackage{soul}
\definecolor{lightblue}{RGB}{185,210,248}
\sethlcolor{lightblue}

\begin{document}

\title{Simulation of many-body dynamics using Rydberg excitons}

\author{Jacob Taylor}
\thanks{These authors contributed equally to this work.}
\affiliation{National Research Council of Canada, 100 Sussex Drive, Ottawa, Ontario K1A 0R6, Canada}
\affiliation{University of Waterloo, 200 University Ave W, Waterloo, Ontario N2L 3G1, Canada}

\author{Sumit Goswami}
\thanks{These authors contributed equally to this work.}
\affiliation{Institute  for  Quantum  Science  and  Technology  and  Department  of  Physics  and  Astronomy, University  of  Calgary, Calgary  T2N  1N4, Alberta, Canada}
\author{Valentin Walther}
\affiliation{ITAMP, Harvard-Smithsonian Center for Astrophysics, Cambridge, Massachusetts 02138, USA}

\author{Michael Spanner}
\affiliation{National Research Council of Canada, 100 Sussex Drive, Ottawa, Ontario K1A 0R6, Canada}
\affiliation{Department of Physics, University of Ottawa, 25 Templeton Street, Ottawa, Ontario, K1N 6N5 Canada}
\author{Christoph Simon}
\affiliation{Institute for Quantum  Science  and  Technology  and  Department  of  Physics  and  Astronomy, University  of  Calgary, Calgary  T2N  1N4, Alberta, Canada}
\author{Khabat Heshami}
\affiliation{National Research Council of Canada, 100 Sussex Drive, Ottawa, Ontario K1A 0R6, Canada}
\affiliation{Department of Physics, University of Ottawa, 25 Templeton Street, Ottawa, Ontario, K1N 6N5 Canada}

\begin{abstract}
The recent observation of high-lying Rydberg states of excitons in semiconductors with relatively high binding energy motivates exploring their applications in quantum nonlinear optics and quantum information processing. Here, we study Rydberg excitation dynamics of a mesoscopic array of excitons to demonstrate its application in simulation of quantum many-body dynamics. We show that the $\mathbb{Z}_2$-ordered phase can be reached using physical parameters available for cuprous oxide (Cu$_2$O) by optimizing driving laser parameters such as duration, intensity, and frequency. In an example, we study the application of our proposed system to solving the Maximum Independent Set (MIS) problem based on the Rydberg blockade effect. 

\end{abstract}

\maketitle
\section{Introduction}

The strong dipole transition between high-lying Rydberg states of atoms is the origin of the long-range interaction between Rydberg atoms~\cite{SaffmanRev, SibalicAdamsBook}. The most notable manifestation of Rydberg-Rydberg interaction is the Rydberg blockade effect where exciting one atom prevents all other atoms in a certain vicinity from being excited to the same Rydberg state~\cite{Urban2009,Gaetan2009}. This feature has inspired several applications in quantum nonlinear optics~\cite{peyronel2012quantum} for development of single photon sources~\cite{muller2013room}, photon-photon gates~\cite{paredes2014all}, and switches~\cite{gorniaczyk2014single}. The application of Rydberg blockade effect has also been explored in generation of entangled states of atoms for atomic clocks~\cite{komar2016quantum} and for entanglement distribution in quantum repeaters~\cite{han2010quantum,PhysRevA.81.052329}. Progress in trapping individual atoms combined with the ability to controllably excite atoms to their Rydberg states is enabling applications in quantum simulation of quantum many-body dynamics~\cite{bernien2017probing,keesling2019quantum} and in generation of many-body entangled states~\cite{omran2019generation}. Rydberg atoms in an optical lattice and the blockade effect provide the opportunity to unlock their applications in finding quantum simulation solutions for mathematical problems such as the Maximum Independent Set (MIS) problem~\cite{pichler2018quantum}.

Bound states of electron-hole pairs (excitons) in semi-conductors are governed by the Coulomb attraction and therefore demonstrate properties similar to those of hydrogen-like atoms. In semiconductors with a relatively high exciton binding energy Rydberg states of excitons can be observed. In~\cite{kazimierczuk2014giant}, the authors observed Rydberg states of excitons in cuprous oxide (Cu$_2$O) and the effect of Rydberg blockade on the absorption spectroscopy outcome. This provides a platform to explore Rydberg excitons in various semiconductors for applications in quantum non-linear optics~\cite{khazali2017single, walther2018giant}, photonic quantum information processing, and in quantum simulation~\cite{ poddubny2019topological}. 

Here, we numerically simulate the many-body dynamics resulting from exciting an array of Rydberg excitons in Cu$_2$O. Our approach enables us to find optimal experimental conditions for quantum simulation of many-body dynamics or to reach exotic quantum states. In particular, we demonstrate that driving laser intensity, frequency and focusing can be controlled to reach the $\mathbb{Z}_2$-ordered quantum many-body state~\cite{bernien2017probing}. In an example, we show the application of our proposed system to finding the solution to the MIS problem based on the Rydberg excitation pattern in a two-dimensional arrangement of Rydberg excitons. Our proposed implementation is inspired by the progress in the Rydberg excitation of individually trapped atoms, offers a path for realization in a solid-state system with potential for integration, and provides a platform to explore many-body quantum dynamics with Rydberg excitons.

\begin{figure}[ht]
\includegraphics*[viewport= 20 230 600 700, width=\linewidth]{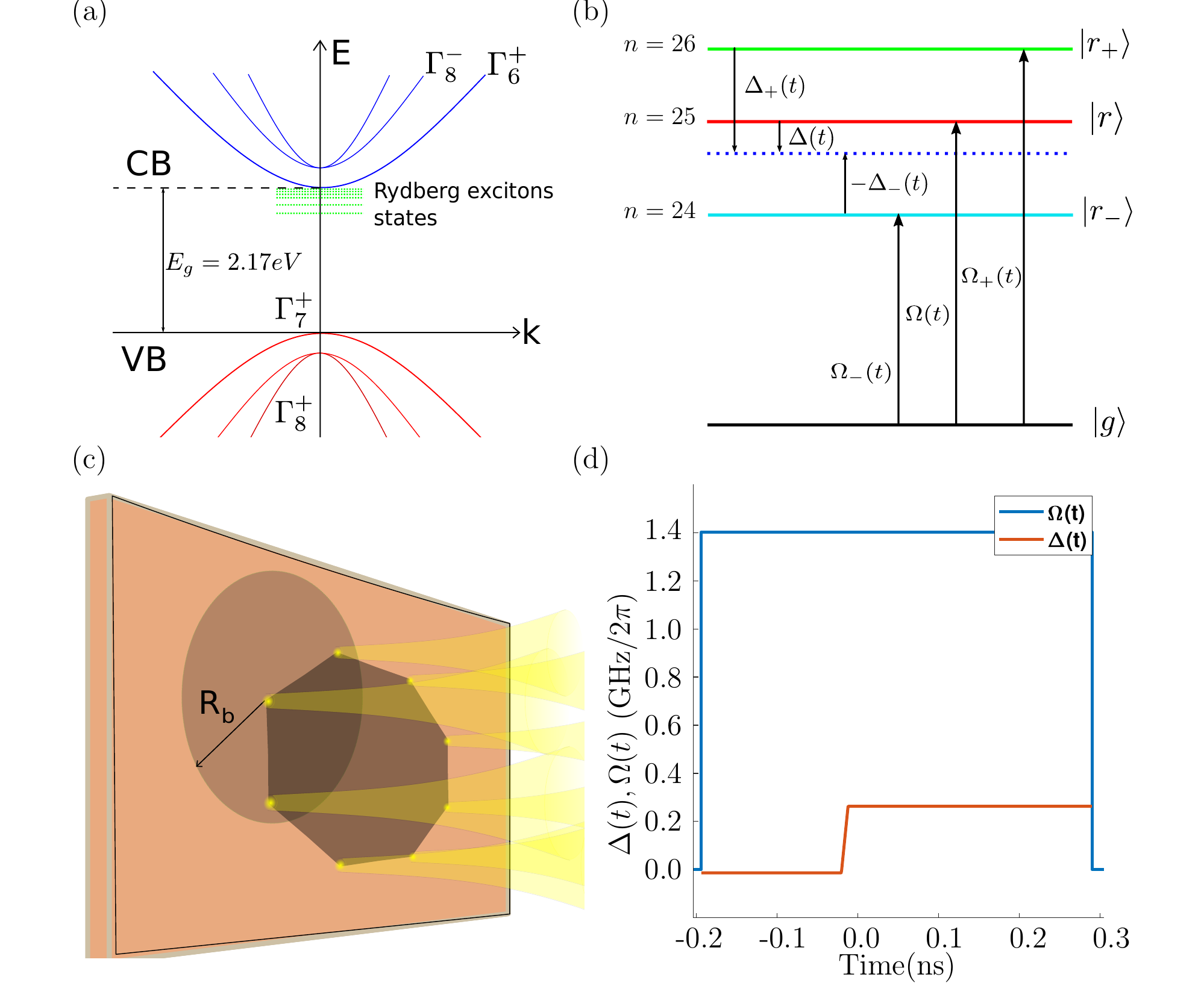}
\caption{(a) Energy band structure of yellow excitons in Cu$_2$O (CB = Conduction Band and VB = Valence Band),  (b) A diagram of the level structure for a single exciton in the case of 3 Rydberg levels, namely the $n$ = 24, $n$ = 25 and $n$ = 26 states; {\it nS} states are accessible through a two-photon excitation mediated by the {\it 2P} state (not shown here). The effective Rabi frequency experienced by different states ($\Omega_k(t)$) varies due to their different transition dipole moments. (c) Schematic diagram showing how multiple exciton sites, in a regular polygon configuration, are created by focused laser excitations at selective positions in a Cu$_2$O micro-crystal. The blockade radius $R_b$ of one exciton is shown which encompasses its nearest neighbours. (d) The shape of Rabi frequency and detuning with time to the $n$ = 25 level is shown here.}
\label{fig1}
\end{figure}

\emph{Scheme--} 
The schematic of cuprous oxide’s (Cu$_2$O) band structure with bound Rydberg exciton states below the conduction band is shown in Fig.~\ref{fig1}(a). These states can be addressed both through single-photon~\cite{kazimierczuk2014giant} and two-photon~\cite{PhysRevB.23.2731} transitions. We aim to address $nS$ states which can be achieved with a two-photon transition via an off-resonant coupling to the $2P$ (intermediate) state; see Fig.~\ref{fig1}(b). The lack of directionality in the Rydberg-Rydberg interaction between $nS$ states enables us to consider various geometries and to eliminate complications arising from angle-dependent interaction between $nP$ states. The excitons could be arranged in a polygon configuration (Fig.~\ref{fig1}(c)) or for that matter in any arbitrary configuration. This is discussed in more detail below. In Fig.~\ref{fig1}(b), we show how neighboring Rydberg states are addressed with an effective Rabi frequency (intermediate state $2P$). As shown in Fig.~\ref{fig1}(c), each excitation site is driven by a focused laser and all sites are exposed to a global secondary exciting laser. Each site is used to define a qubit; $\ket{g}$ and $\ket{r}$ are the states of the qubit where $\ket{g}$ represents lack of an exciton and $\ket{r}$ an exciton in the Rydberg level.

As shown in Fig.~\ref{fig1}(b), each exciton is treated as a multi-level system being addressed with an effective Rabi frequency resulting from two driving lasers. This is in particular important for Cu$_2$O where Rydberg states have relatively small energy spacing. At each site, the lasers are focused in a tiny spot on a very thin crystal such that a small crystal volume - smaller than the Rydberg Blockade volume - is illuminated prohibiting multiple excitations in one site. Fig.~\ref{fig1}(d) shows an example profile for the effective Rabi frequency and two-photon detuning pertaining to our numerical results that are discussed below.

The paper is organized as follows. After the introduction in Sec I, Sec II presents the model for our quantum simulation and discusses the results for a polygon configuration of excitons. We discuss the potential practical application of our scheme to solve MIS problems in Sec III. Sec IV depicts the road ahead for the implementation of our model. There a comparison is made with the atomic case with directions for future improvement, and challenges like losses and various other issues associated with implementation are discussed. We conclude in Sec V by summarizing our results.

\begin{figure*}[ht]
\includegraphics*[viewport = 0 420 600 655, width=\textwidth]{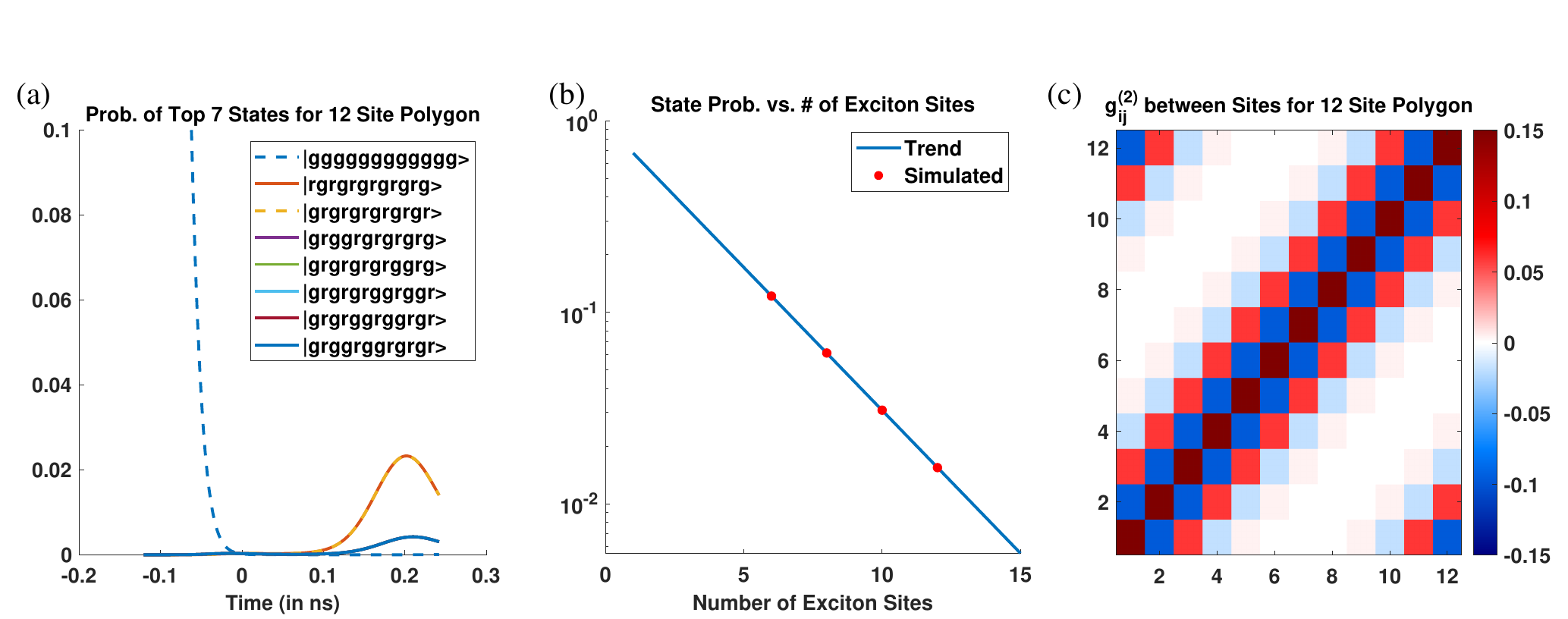}
\caption{(a) Probabilities are plotted with time for the 7 most probable ensemble states and the initial state with no excitons, $|gggggggggggg\rangle$ (dashed blue line), for 12 exciton sites in a polygon configuration. For 12 sites the most probable states, $|rgrgrgrgrgrg\rangle$ and $|grgrgrgrgrgr\rangle$, are equally likely. (b) Probabilities of achieving the objective state(s) for simulations with different number of exciton sites are shown. The scaling can be modelled with a exponential decay as shown with the blue line. (c) Correlation function $g^{(2)}_{ij}$, defined in Eq.~(\ref{g2}) between two excitons at specific positions ($i$ and $j$), for 12 excitons with periodic boundary conditions. Exciton sites neighboring each other, due to the blockade effect, are mostly anti-correlated. The next nearest neighbour sites are then mostly correlated. }
\label{fig2}
\end{figure*}

\section{Simulation}
We study the many-body Rydberg excitation dynamics under the influence of population dynamics and Rydberg-Rydberg interaction. An exciton in the Rydberg level has a large dipole moment which prohibits an excitation within a certain distance, called blockade radius ($R_b$). In our scheme (see Fig.~\ref{fig1}(c)), the blockade radius $R_b$ contains the neighbouring exciton site but not the next nearest-neighbour site. Hence, the Rydberg blockade effect blocks the two nearest neighbors from being excited together creating the $\mathbb{Z}_2$-ordered phase, i.e. states of the form $|rgrgrgr...\rangle$ or $|grgrgrg...\rangle$. Similar to these $\mathbb{Z}_2$-ordered states, our general objective states are the states with the maximum number of excitations that can occur without any additional excitations within the blockade radius of a Rydberg exciton. Our numerical treatment of the many-body dynamics aims to find the optimum pulse energy and detuning to reach these objective states with the highest probability of success. To afford numerical simulation of several multi-level systems we do not include the effect of excitation decay. To circumvent this limitation, we operate at a fast excitation regime where the total population dynamics remains faster than the expected lifetime of the Rydberg states; see below for a more detailed discussion.

The excitons are detected by applying a detection laser (de-excitation) pulse to take them to the $2P$ state and then collecting the photons emitted from the $2P$ exciton decay. Collection fibers and detectors can be placed behind every site to collect emitted photons. However, this detection scheme is rather inefficient as the probability of collecting photons in one direction from the excited state is small. The detection scheme and alternate approaches to increase the detection efficiency dramatically are discussed in detail in Section \ref{sub_sec_det}.

In our proposal based on Cu$_2$O, excitons in $nS$ states, can only be excited via two-photon excitation in a ladder configuration through the intermediate $P$ states, owing to the selection rules. Off-resonant coupling to the transition that promotes an exciton to the $2P$ state provides the flexibility to operate at various wavelengths. The two-photon excitation scheme~\cite{PhysRevLett.125.173601, PhysRevB.96.115207} off-resonantly through an intermediate state ({\it e.g.} the $2P$ state) can help circumvent the underlying phonon-assisted absorption~\cite{kazimierczuk2014giant}. This approach is expected to enable sharp two-photon transitions with optical depth and Rydberg blockade properties comparable to that of cold atomic ensembles~\cite{PhysRevLett.125.173601}. For the single-photon detuning of $\delta$ with respect to the $\ket{2P}$, the effective Rabi frequency representing the coupling to the $\ket{nS}$ is given by $\Omega = \Omega_1 \Omega_2/ \delta$, where $\Omega_{1,2}=\langle E_{1,2}\cdot d_{g2P,2PnS}\rangle/2\hbar$. Here $E_{1,2}$ are electric field associated with the two excitation fields and $d_{ij}$ is the transition dipole moment for $\ket{g}\rightarrow \ket{2P}$ and $\ket{2P}\rightarrow \ket{nS}$ transitions. The ground state here represents a lack of exciton and the mode is determined by that of the $E_1$. For large $\delta$ (i.e. $\delta >> \Omega_{1},\Omega_{2}$), we can describe the two-photon transition with the effective Rabi frequency, $\Omega$.  Such two-photon excitation was long used to excite $nS$ Rydberg excitons~\citep{uihlein1981investigation} with optical and infrared lasers. A degenerate two-photon excitation with a laser at 1eV could simplify the experimental condition.
Resonant coupling to the intermediate state is also a possibility based on electromagnetically induced transparency as explored in \citep{walther2018giant, PhysRevLett.125.173601} - using an optical and another THz laser - to show enhanced non-linearity through Rydberg excitons in an optical cavity. 

Contrary to the Rydberg atoms, in Cu$_2$O excitons, the separation between Rydberg levels is not much greater than the linewidth of these levels. The smaller ratio of line separation and linewidth is a major challenge for effective excitation of  Cu$_2$O excitons to a single Rydberg level; which can be important for quantum information processing applications. While trying to excite the intended excited state (say, $n$ = 25) we may inadvertently excite adjacent levels too ($n$ = 24, $n$ = 26 etc). This significantly decreases the probability of achieving the objective state(s) in the Cu$_2$O exciton system compared to the atomic system, as discussed later. Hence, this effect is needed to be included for an effective simulation of the quantum dynamics. For this purpose, we model each Cu$_2$O exciton as a 4-level system, 1 ground and 3 Rydberg levels system as shown in Fig.~\ref{fig1}(b). We aim to excite the $n$ = 25 level. For this high lying Rydberg level ($n$ = 25) the energy gaps to the adjacent Rydberg levels are relatively small, $2\pi\times 2.81$~GHz above and $2\pi\times 3.18$~GHz below \cite{kazimierczuk2014giant} while the linewidth of $n$ = 25 level is around $2\pi \times 0.102$~GHz.

In our model, each exciton is driven by a laser with $k^{th}$ Rydberg state coupled by effective Rabi frequency $\Omega_k(t)$ and detuning $\Delta_k(t)$.  The interaction potential between two exitons at $i^{th}$ and $j^{th}$ position is modeled by the principal number dependent van der Waals interaction $V_{k_{ij}}=\frac{C_k}{R_{ij}^6}$, where $R_{ij}$ is the distance between the two excitons and $C_k$ is the interaction coefficient corresponding to the $k^{th}$ Rydberg level. The Hamiltonian governing such a system is given by
\begin{equation}
    \frac{H}{\hbar}= \sum_{k,i}  \frac{\Omega_k(t)}{2} (\sigma_i^k)-\sum_{k,i} \Delta_k(t) n_i^k +\sum_{k,i>j} V_{k_{ij}} n_i^k n_j^k,
    \label{hamiltonian}
\end{equation}
where  $\ket{g_{i}^{\sp{}}}$ and $\ket{r_{i}^{k}}$  represent the the lack of an exciton and an exciton in kth Rydberg level respectively at the ith position (site) in the arrangement. $\sigma_i^k=\ket{r_{i}^{k}} \bra{g_i^{\sp{}}}+\ket{g_i^{\sp{}}} \bra{r_{i}^{k}}$, $n_i^k=\ket{r_{i}^{k}} \bra{r_{i}^{k}}$, and $V_{k_{ij}}$ is the interaction potential between excitons $i$ and $j$. As it can be seen in Eq.~\ref{hamiltonian}, the interaction between different Rydberg states such as $24S-25S$ is not included in this model. Such cross-interaction terms brings a breadth of possibilities specially in study of wave-packet dynamics in Rydberg excitons where a superposition of multiple Rydberg states of the same exciton is present~\cite{heckotter2020asymmetric}. Here, we minimize excitation to states other than the targeted $25S$ state numerically to eliminate the effect of these cross-interaction terms.

Using a reference Rydberg state, the detuning with respect to the other adjacent states, $\Delta_{k'-1}$ and $\Delta_{k'+1}$, are calculated from the energy gaps \cite{kazimierczuk2014giant}. Thus given $k'$ as a reference level, $R_y$ the Rydberg constant and $\delta_p$ the quantum defect, $\Delta_k(t)$ is calculated as
$\Delta_k(t)=\Delta_{k'}(t)+\frac{R_y}{(k'-\delta_p)^2}-\frac{R_y}{(k-\delta_p)^2}.$
Similarly we calculate the Rabi frequency for each Rydberg level using a set of Rabi frequencies and the relative electric dipole moments ($\mu_k$), given a reference $k'$ then $\Omega_k(t)=\sqrt{\frac{\mu_k}{\mu_{k'}}}\Omega_{k'}(t)$.
In our case the reference $k'$ is the $k'$=25 state, for all parameters. We propose the same square pulse for all the lasers, so the Rabi frequency is constant in time and space ($\Omega_k'=\Omega$) with $\Omega=2 \pi \times 1.404~$GHz, while $\Delta(t)$ is time dependent and is iteratively optimized to maximize the probability of reaching a selected objective state(s). In particular $\Delta(t)=at^3+bt+c$ is selected where $a$, $b$, $c$ are constants optimized for a small system size (such as 5) which is then used for the computation of larger site systems. This $\Delta(t)$ is also truncated from above/below by $\Delta_\text{max}$, $\Delta_\text{min}$ respectively, are selected as to minimize the excitation of neighboring states. The shape of $\Omega(t)$ and $\Delta(t)$ used for the simulations are shown in Fig.~\ref{fig1}(d).

We choose the distance between two excitons (i.e. the nearest neighbor distance) as a fraction of the blockade radius. The blockade radius is the distance between the exciton sites such that  $V_{ij}=\Omega$, or more directly $R_b=\sqrt[6]{\frac{C_6}{\Omega}}$. The distance was chosen to be about $0.958\times R_b= 2.72\mu m$ \cite{walther2018interactions}. By having the exciton sites separated by this distance, nearest neighbors have $V_{k_{ij}}>>\Omega_k$, while second nearest neighbor have $\Omega_k>>V_{k_{ij}}$. This effect, referred to as Rydberg blockade, blocks nearest neighbors from becoming excited. This was done with the goal of obtaining an ordered objective state(s). 

Our objective state(s) refers to the state(s) with the maximum number of excitations that can occur without any nearest neighbors being excited for any arrangement of the exciton sites. Hence, in our simulation where 12 exciton sites are arranged in a regular polygon, the objective states would consist of two states - $|rgrgrgrgrgrg\rangle$ and $|grgrgrgrgrgr\rangle$. When the sites in a polygon are excited by lasers a superposition of these states should ideally be the final state of the system with high probability. We numerically simulated this dynamics. The state probabilities are plotted over time in Fig.~\ref{fig2}(a). The objective states for the 12-site polygon are the most probable states. In Fig.~\ref{fig2}(a), we show the population dynamics for generation of the objective states with any of the three neighboring Rydberg states excited. As we aim to minimize Rydberg excitation in neighboring states this is relatively close to dynamics of generating the objective state with a specific Rydberg state denoted as $|r_1gr_1...gr_1g\rangle$ and $|gr_1g...r_1gr_1\rangle$.  

The maximum probability of the objective state(s) ($|r_1gr_1...gr_1g\rangle$ and $|gr_1g...r_1gr_1\rangle$) decreases with more exciton sites, as shown in Fig.~\ref{fig2}(b). For a set of 6, 8, 10, and 12 exciton sites probabilities to achieve the objective state are 0.1214, 0.06127, 0.03082 and 0.01545, respectively. Here, it should be noted although the shape of $\Delta(t)$ with time is optimized to get the highest probability, the same $\Delta(t)$ profile was used for all the different exciton sites. This is to ensure scalability of our approach so that the optimization results can be applied to higher number of excitons where solving the population dynamics become intractable.

The probabilities in Fig.~\ref{fig2}(b), similar to the system of individually trapped atoms (see Fig.~4 in Ref.~\cite{bernien2017probing}), shows a clear exponential trend. Extrapolating this trend we calculate that a polygon of 50 exciton sites would be expected to attain the objective state with a probability of around $P = (3.30 \pm 0.07)\times 10^{-8}$. This estimate is motivated by the experimental demonstration of the many-body ordered phase in an array of individually trapped atoms~\cite{bernien2017probing} and will allow us to gain perspective over our numerical results.

Full quantum state tomography remains very challenging as it requires a high number of complementary measurements. An alternative and very important metric to measure the ordered objective state is by measuring the second order atom-atom correlation function defined as - 
\begin{equation}
    g^{(2)}_{ij}= \langle n_in_j \rangle - \langle n_i \rangle \langle n_j \rangle,
    \label{g2}
\end{equation}
where the average ($<...>$) is taken over many repetitions. The blockade effect is apparent within Fig.~\ref{fig2}(c) where a positive value implies correlation while negative value implies anti-correlation. The anti-correlation between the neighbouring sites and the correlation between next-nearest neighbours shows a critical signature of the $\mathbb{Z}_2$-ordered phase induced by Rydberg blockade between the nearest neighbours. We also see correlation in top-left or bottom-right corner of Fig.~\ref{fig2}(c) because the arrangement here is periodic - e.g. site 1 is also the neighbour of site 12 and hence they are correlated. 

Although the objective states we expect ($|rgrgrgrgrgrg\rangle$ and $|grgrgrgrgrgr\rangle$) - described as the $\mathbb{Z}_2$ order in \cite{bernien2017probing} - has complete correlations, these only occurs with probability $2\times0.015$ while other states with partial or no correlation also occur (Fig. \ref{fig2}(a)). As the correlation function, $g^{(2)}$, is averaged over many runs of the simulation it dies down quickly after a few sites. This is expected and similar results occurred in trapped atom simulation experiments \citep{bernien2017probing}. However, this makes $g^{(2)}$ an interesting and accessible experimental signature as this correlation can be seen even in the presence of significant loss.

\section{Application}
\begin{figure*}[ht]
\includegraphics*[viewport = 0 440 600 655, width=\textwidth]{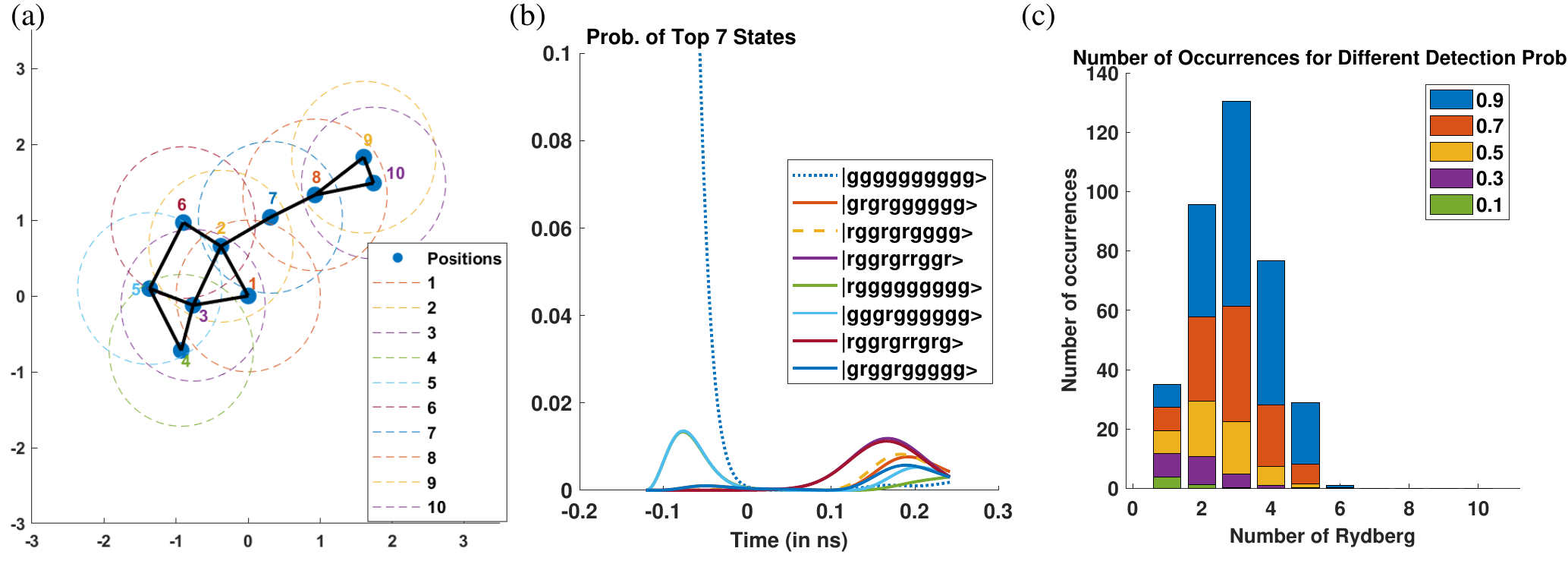}
\caption{ (a) A 10 site graph with sites arranged arbitrarily in 2D to solve a Maximum Independent Set (MIS) problem using the excitons. The blockade radius of each exciton is shown by a dotted circle around it and if the separation between two excitons is less than the blockade radius they are connected by an edge creating the graph. (b) Probabilities over time of achieving different states for the 10 site graph. The probability of achieving the MIS objective states was found to be 0.01187 and 0.01124, with a total probability of 0.02311. (c) The number of occurrences for different amounts of Rydberg excitons, based off 2000 runs. The histogram is shown for expected number of detections given non-unitary detection probabilities of 0.9, 0.7, 0.5, 0.3 and 0.1. }
\label{fig3}
\end{figure*}

\emph{Maximum Independent Set (MIS) problem} aims to find the largest subset of nodes in a graph where no two nodes are adjacent (linked by an edge). It has been shown that solution to NP-complete problems such as the MIS problem on planar graphs can be mapped onto the ground state of a multi-body Rydberg system with proper arrangement of Rydberg atoms (or in our case excitons); see~\cite{pichler2018computational}. This ground state is our objective state with maximum number of Rydberg excitations such that no two excitations occur within the Rydberg blockade radius. If there are two excitations within the blockade radius the total energy goes up and it is no more the ground state. The Rydberg blockade effect combined with a proper positioning of Rydberg excitons will allow us to naturally reach a solution for the MIS problem. Finding the correct position for each Rydberg exciton is an exponentially challenging task for large graphs. Below we treat an example with a graph of 10 nodes as shown in Fig.~\ref{fig3}(a).

The polygon scheme described before was a trivial MIS configuration, with the objective state easy to find theoretically. Although we observed that the probability of achieving the objective state is not very high in one repetition of the experiment - $P = 2\times0.015$ for 12 sites in a polygon case -  that wouldn't eliminate our ability to find a solution as a solution is easy to check in polynomial time. Given a solution one simply needs to check if two of the excitations are within blockade radius - if not it is a 'valid' solution and one looks for the valid solution with maximum number of excitations. All that would be required is enough experimental runs to have a high enough confidence there is no 'valid' larger set of excitations.

We depict such an MIS case in Fig.~\ref{fig3}. Part (a) shows the ten exciton configuration arbitrarily placed in 2D with the solid lines showing the associated planar graph. The evolution of different possible final states of the exciton ensemble are shown in Fig.~\ref{fig3}(b). The objective states (which are called 'ground states' in \cite{pichler2018computational}) with the maximum possible five excitons each, $|rggrgrrggr\rangle$ and $|rggrgrrgrg\rangle$, have the highest probabilities,  0.01187 and 0.01124, with a total probability of 0.02311. The number of Rydberg excitation in the final state is plotted as an histogram in Fig.~\ref{fig3}(c) over 2000 runs. We see a finite number of occurrences for five excitons but six excitons also occur in small number (at detection efficiency 0.9). Creation of six excitons is an error due to the presence of other nearby Rydberg states (see, Fig.~\ref{fig1}(b)). It can be easily checked that no state associated with six excitation is 'valid', as defined earlier. So, we conclude the five excitation states are our objective (ground) states. Fig.~\ref{fig3}(c) shows the effect of finite detection efficiencies too, as Rydberg excitation numbers are indicated in different color for different detection efficiencies. Naturally, for lower detection efficiencies we expect to detect smaller number of excitons as more excitons in the final state remain undetected. Low detection efficiency limits us in finding the objective state with maximum number of excitons; however, as it can be seen in Fig.~\ref{fig3} even for a detection efficiency of 0.7, 2000 runs (attempts) is enough to find the solution to the associted MIS problem. The objective state can be found at even lower detection efficiencies by increasing the number of runs.

\section{Implementation}

\subsection{Comparison with atoms and improvements} 
In the following, a comparative analysis is drawn between Rydberg atoms and Rydberg excitons (in Cu$_2$O) for their application as a quantum simulator. This will allow us to put our results in perspective and motivate experimental efforts based on Rydberg excitons. Significant progress in trapping individual atoms and well-known Rydberg properties and excitation strategies puts the atomic systems at a significant advantage. However, the relatively slow trapping, excitation, and detection steps (in order of milliseconds) can become a bottleneck in applications that require sampling over many outcomes. On the other hand, for the fast exciton system trapping is not required which will offer simplicity and operation time in the order of nanoseconds. 

The closely placed absorption lines of Rydberg excitons (in Cu$_2$O) prohibit effective excitation of Rydberg states resulting in low probability of success in achieving the objective state. The probabilities achieved in our Rydberg exciton simulation are much lower than what is achievable with trapped atoms \cite{bernien2017probing}. In Cu$_2$O excitons, the radiative linewidth of $n$ = 25 state is $2\pi *0.102$~GHz as compared to the $2\pi \times 2.81$~GHz line separation between $n$ = 25 and $n$ = 26 levels. The linewidth is further broadened due to power broadening effects. The small line-separation limits the range of $\Omega$ that can be used. To generate the objective state one must prevent excitations into higher/lower Rydberg levels ($ n = 24 , 26$ in our model). It is therefore desirable to have $\Delta_k>\Omega_k$ for $k \neq 25$. In sharp contrast, this condition is very easily satisfied in the atomic system where the atomic linewidth is millions of times smaller than the line separation. Fortunately, this large linewidth and hence the small lifetime of excitons is not without benefits, as it also allows many iterations of the experiment to be performed orders of magnitude faster than individually trapped Rydberg atoms. Operating on orders of nanoseconds instead of 100 $\mu s$, as in the atoms. This allows one to do 100,000s of times more repetitions in the same time, increasing the number of successful attempts in generating the objective state. This in turn increases the probability of a successful detection and hence identification of the objective state.

To compare these two different systems quantitatively we ask the following question - for how much time $T$ must the experiment be run to achieve a particular probability (say $99\%$) of identifying the objective state of the simulation with {\it n} number of sites?

Detection probability is taken as $99\%$ for each atom. It is similarly high for excitons, which is justified for the alternative detection scheme discussed in Section \ref{sub_sec_det}. For the excitons we use a more conservative $90\%$ detection probability here combining the effect of decay loss too. This is discussed and justified in detail in Section \ref{decay_loss}. 

A single experimental run time takes around 250 ms for the atoms \citep{bernien2017probing} and estimated to be around 5 ns for the exciton. Assuming the above parameters, for the 50 site case, to get a $99\%$ chance of the correct state occurring at least once, about 1.94 billion runs need to be made for the exciton case. This can be achieved in around 9.72 seconds. While for the atomic case only 842 runs are needed. However, due to the larger individual run times this would take around 210 seconds.

A smaller time for single experimental run lowers $T$ by a fixed factor, while a better scaling of success probability with $n$ (data in Fig. 4a of \citep{bernien2017probing} for atoms and in Fig~\ref{fig2}(b) here for excitons) enhances $T$ slower for higher $n$. Hence, the atomic system perform favorably for higher value of n (more number sites) while exciton system (in Cu$_2$O) is faster for smaller $n$ values. 

It can be concluded from the above analysis that to scale to a really large number of sites, in the long term, one would actually need to improve line-separation vs linewidth ratio which differentiated the excitons from the atoms. This has been a problem in Cu$_2$O system as other unwanted nearby states get excited which resulted in ineffective Rydberg excitation and correspondingly worse success probability scaling. In the Cu$_2$O system, the small lifetime and the corresponding large linewidth of the Rydberg excitonic states caused this. Optical cavities could be used to enhance the coupling efficiency or to reduce the decay rate in such a system. This deserves a dedicated study to find optimal conditions for quantum nonlinear optics with Rydberg excitons. Another possible solution to this problem may be to use engineered micro structures to enhance the lifetime. Such a system has recently been experimentally demonstrated in the context of exciton condensation, in MoSe$_2$–WSe$_2$ double layer material where the inter-layer excitons demonstrate long lifetimes and maintain the large binding energy which is required to observe and address Rydberg states~\cite{wang2019evidence}. Optical addressing and lifetime measurement of excitons in such structures have recently been studied in~\cite{montblanch2020confinement} with exciton lifetime of 0.2~ms. Although such long lifetimes may not actually be needed. Possibly a lifetime close to a microsecond would be enough as that would already mean a line separation vs linewidth ratio above 1000. However, there would be complications for long lifetimes as we would need trapping to keep the excitons in one place which would otherwise move due to thermal motion, discussed in details in Section \ref{sub_sec_other} . However, there is an important difference with the trapped atoms. Rydberg excitons can be trapped mechanically using strain traps \citep{kruger2018waveguides}. Such mechanical traps can be part of fabrication or prepared with the system. This would save precious trap preparation time for each iteration of the experiment, as needed for optical traps containing Rydberg atoms. 

\subsection{Decay loss}\label{decay_loss}

There are two principal sources of loss in our system - decay loss and detection loss. Detection loss and ways to avoid it are presented in detail in Section \ref{sub_sec_det}. The decay loss is caused by the decay of Rydberg states during the simulation. The objective state decays at a rate proportional to the number of excitons present. Thus given a fully connected ensemble in a regular polygon arrangement, the worst possible decay rate is proportional to 1/2 the size of the ensemble. For fewer Rydberg excitations, in partially connected ensembles, the effect of decay is even less. The low probability of Rydberg excitation at early part of the dynamics (see Fig.~\ref{fig2}(a)) and our fast excitation timescales with respect to the lifetime allow us to reliably remove excitation decay from our numerical simulations for small number of sites. This helps to put computational resources towards treating higher number of qubits. The objective state has a significant probability of being excited only starting from 0.1 ns (see Fig.~\ref{fig2}(a)). As the simulation ends at 0.24 ns, considering $\sim$ 1 ns lifetime of the state, an individual exciton has a maximum of 14$\%$ chance of decay during simulation. However, this is only if an exciton is already excited by 0.1 ns, i.e. when the excitation probability is just picking up (as in Fig. \ref{fig2}(a)). On average, excitons would be excited much later and hence the decay probability would also be smaller. As a crude estimation, it can be considered around half of that ($\sim 7 \%$).

Even then, for a large number of sites, exciton decay can no longer be neglected for theoretical simulation of the final state. However, for the practical purpose of detecting the objective state in an experiment, the exciton decay loss plays an equivalent role to the detection loss. The two losses manifest themselves differently - decay loss produces incorrect final states of simulation and detection loss inhibits the correct detection of the final state. But they both obstruct finding the objective state. With enough number of experimental runs one can find the objective state in the presence of each of the two losses. Hence, the maximum possible loss would simply be a combination of the two losses and the two losses can be treated equivalently for the purpose of finding the objective state; our prime goal in this proposal. Hence, even if we extrapolate our results for much larger number of sites (as we did for 50 sites using Fig.~\ref{fig2}(b)) without considering the effect of exciton decay, we would still find the correct minimum probability of detecting the objective state(s) as long as we combine the decay loss with the corresponding detection loss.

\subsection{Detection}\label{sub_sec_det}

\begin{figure}[ht]
\includegraphics*[width=\linewidth]{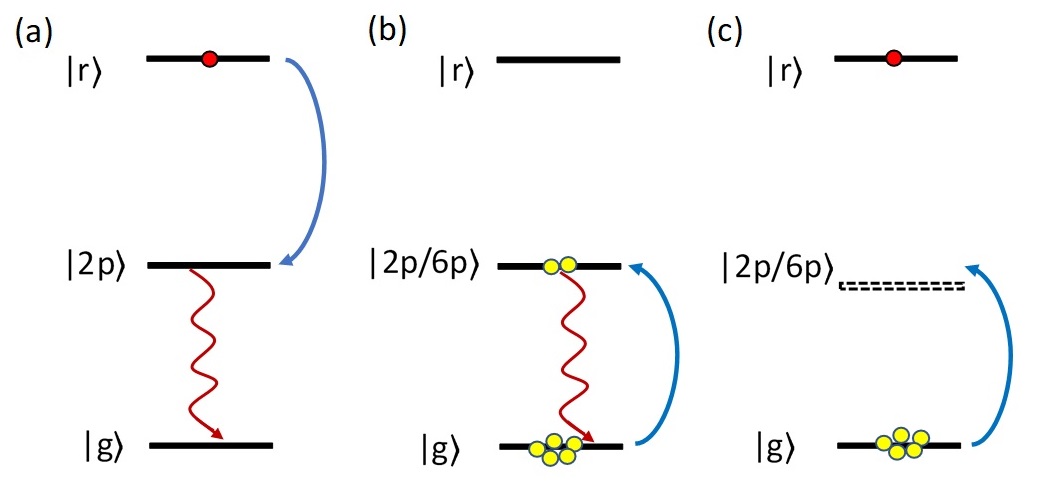}
\caption{The two different detection schemes are shown. In the original scheme (a) Rydberg excitons in $|r\rangle$ = $|25S\rangle$ state are detected directly by first transferring it to $|2P\rangle$ state and then detecting the emitted photon. In the alternate scheme the presence of the $|r\rangle$ state is detected indirectly, as a non-demolition detection, through many-body fluorescence detection. Detection laser pulse is applied to create low lying short lived $P$ excitons which decays by emitting photon that gets detected. The sequence is later repeated to detect a large number of photons from each site. Moreover, many $P$ excitons are created at each site in each iteration as they are small in size compared to the $|r\rangle$ exciton. Such $P$ excitons are created if no $|r\rangle$ exciton is present (b) but in presence of a $|r\rangle$ exciton creation of $P$ excitons gets hindered due to the blockade effect (c).}
\label{fig_det}
\end{figure}

Faithful detection of the final state is an important requirement of our scheme. In the original detection scheme, we try to detect the $|r\rangle = |25S\rangle$ state by making it decay and detecting emitted photon. However, $|25s\rangle$ state can only decay by two photon emission. Hence, as shown in Fig~\ref{fig_det} (a) we transfer the $|25S\rangle$ state to $|2P\rangle$ state by a THz detection laser and detect the photon emitted from the $|2P\rangle$ state. This detection scheme completely depends on the ability to detect the one photon emitted from the $|2P\rangle$ state. Even if one reaches a very high combined collection and detection efficiency (say, around 90$\%$) of detecting the single photon or equivalently a single Rydberg excitation, the probability of detecting the final state decreases exponentially with larger number of excitons. Moreover, the large size of Rydberg excitons ($\sim$ 1$\mu m$) compared to their blockade radius ($\sim$ 2.82 $\mu m$) means Rydberg excitation lasers (used for simulation) must be focused to a small area in the middle of the crystal resulting in excitation in a single exciton compared to a collective excitation over multiple excitons. Hence, there would be no preferred directions to the emitted photon in contrast to collective emission. This random direction of emission, coupled with limited coupling efficiency with the collecting fiber and finite detector efficiency lowers the photon detection efficiency. One way to enhance efficiency is using an optical cavity which would result in a directional emission even from a single exciton. Cavity would also enhance the coupling strength of the relevant exction line while not enhancing the other off-resonant lines which would suppress the off-resonant excitations. However, fabricating such a cavity would also be challenging as it needs to have a quality factor in hundreds of thousands. In this context, it should be noted - as we have stressed before - low detection efficiency doesn't affect $g_2$ correlation function considerably as it is averaged over. 

To solve this problem, in the following we propose an alternate detection scheme which drastically improves the detection efficiency. Instead of directly detecting the Rydberg excitation by making it decay, here we detect the presence of the Rydberg state $|r\rangle$ indirectly, as a non-demolition detection. The scheme is based on cyclic transitions as shown in Fig~\ref{fig_det}(b)-(c). Detection lasers pulses are applied at all sites to create low lying Rydberg excitons (say, 2P or 6P state). We try to detect the emitted photons and the sequence is repeated.

In the absence of a Rydberg excitation (i.e. at the sites in $|g\rangle$ state) $|2P/6P\rangle$ excitons would be created and decay by emitting photons (Fig~\ref{fig_det} (c)). While in the presence of a Rydberg excitation (i.e. at the sites in $|r\rangle$ state) ideally no $|2P/6P\rangle$ excitons should be excited due to the blockade effect (Fig~\ref{fig_det}(b)). Such asymmetric Rydberg interaction exists and has been observed between 6P and 17P states of Cu2O excitons. This was prominently seen in Fig. 3(e) of \cite{kazimierczuk2014giant}. We expect a similar effect between $|25S\rangle$-$|2P/6P\rangle$ excitons will allow us to implement our proposed detection strategy. This would result in reduced excitation probability at a site in $|r\rangle$ state compared to one in $|g\rangle$ state. Hence $|g\rangle$ and $|r\rangle$ states can be easily distinguished if a large enough number of photons can be collected. The alternate detection scheme constitutes a non-demolition measurement of the Rydberg state. Although, it does constitute a measurement, e.g. destroying the coherence between two degenerate $\mathbb{Z}_2$ eigenstates. P excitons are excited resonantly here, contrary to the two-photon excitation of S excitons, resulting in a phonon-assisted background absorption. However, the background is not significant compared to the transition line. It is complicated to include the exciton-phonon interaction in calculations. Hence, we neglect the effect of the background absorption in our calculation.

The relatively small energy gap between Rydberg states and the asymmetric Rydberg-Rydberg interaction raises a question whether all possible cross-interaction terms ($n’S-nS$ for $n’\neq n$) should also be included in Eq.~\ref{hamiltonian}. In this work, we minimized excitation in neighboring Rydberg states (where $k\neq 25$) to eliminate the effect of such cross-interaction terms on the Rydberg population dynamics. For broadband addressing of Rydberg states and study of Rydberg exciton wavepacket dynamics, it is essential to include all potentially relevant cross-interaction terms. This novel complication introduces an opportunity to develop novel applications of Rydberg excitons in quantum information processing.

Note that the alternative detection scheme doesn't require a THz laser, only a yellow optical laser is needed. Also note that it is tricky to talk about $|g\rangle$ state and $|r\rangle$ state anymore as we are creating multiple excitons in the low lying $p$ state. We discussed in introduction how multiple excitons can break our approximation of two level system. However, please observe that no additional $|25S\rangle$ exciton is excited due to the blockade as before. So, for simplicity we will go on with the notation by redefining $|g\rangle$ and $|r\rangle$ state as absence and presence of $|25S\rangle$ exciton only.

The principal advantage of the alternative detection scheme is that the number of photons collected from one site is huge as compared to maximum one in the original detection scheme. This is due to three reasons. Firstly, we perform a cyclic transition by repeatedly creating and detecting the $|2P/6P\rangle$ excitons. This can be repeated many times within the lifetime of the $|25S\rangle$ exciton as the lifetime of $|2P/6P\rangle$ excitons are much smaller. Secondly, as the size of $|2P/6P\rangle$ excitons are much smaller than the $|25S\rangle$ exciton many $|2P/6P\rangle$ excitons can be excited simultaneously which would result in detection of multiple photons each time. Thirdly, due to this ensemble excitation of $|2P/6P\rangle$ excitons the correspondingly emitted photons would be directional enhancing the photon collection efficiency. Please note that we assumed that exciton linewidths to be radiative here, based on current evidences \citep{kazimierczuk2014giant}.  Although the contribution of phonon-mediated line broadening in excitons is not completely clear, especially to the lower $n$ values. This may adversely affect the detection efficiency.

We quantify the number of emitted photons from a site in $|g\rangle$ state to show that indeed a very large number of photons are emitted. Lifetime of $n^{th}$ state is proportional to $n^3$. So, a transition to lower $n$ levels can be repeated many times inside the lifetime of the $|n=25\rangle$ state ($\sim$ 1953 times to $n$=2 and $\sim$ 72 times to $n$=6 state). Moreover, the size of lower lying excitons are much smaller, given by spherical volumes with radius $\langle r_n\rangle = \frac{1}{2} a_B (3n^2 - l(l+1))$, where Bohr radius $a_B$ = 1.11 nm and for s excitons $l$ = 0 while for p exciton states $l$ = 1. This gives a radius of $1040.6$ nm for 25$S$ state while only $5.6$ nm and $58.8$ nm for 2$P$ and 6$P$ states respectively. Assuming we wait for three lifetimes (of $P$ state) to let all the excitons emit, the combined effect of both factors would result in emission of a staggering number of photons, $4.3 \times 10^9$ and $1.3 \times 10^5$ for 2$S$ and 6$S$ states respectively. Here, we have considered the exciton radius itself instead of the blockade radius because blockade radius would depend on the choice of detection pulse Rabi frequncy which is flexible. Also, the value of blockade radius generally stays in the same order as the exciton radius.

We estimated the mean photon number above. But to distinguish between the $|g\rangle$ and $|r\rangle$ states by measuring photon number, we need to know the photon number distribution too. We assume the probability to successfully detect a particular exciton by photon detection is $P$ - i.e. combining probabilities of photon emission within the detection time window, within the spatial collection angle and successful detection. If a total $N$ number of excitons are attempted to be detected (in time and space), their distribution is binomial resulting in mean photon detected $\overline{x} = Np$ and variance $\sigma^2 = Np(1-p)$. As binomial distributions are almost Gaussian for high $N$, to have a vanishing overlap (less than $1\%$) between the two distribution ($|g\rangle$ and $|r\rangle$) the mean separation needs to be more than $6 \sigma$. As $\sigma << \overline{x}$ in our case we can assume both $\sigma$ and $\overline{x}$ to be almost same for both distribution. A  6$\sigma$ separation would imply, detection fidelity of 99.9$\%$. We estimate below how much mean separation (i.e. 6$\sigma$) is needed as fraction of mean to achieve this. In our case, we have  $6\sigma/\overline{x}  = 6\sqrt{\frac{1-p}{Np}}$. Using $N$ from above and assuming a photon collection probability of $p$ = 0.5 we would have $6\sigma/\overline{x}$ of 0.000091 and 0.017 for $n$ =2 and $n$ = 6 cases respectively. Hence, to distinguish $|g\rangle$ and $|r\rangle$ states with 99.9$\%$ fidelity we need only 0.0091$\%$ (or 1.7$\%$) less photon emitted in case of $|r\rangle$ state than in case of $|g\rangle$ state while using $n$ = 2 ($n$ = 6) state. This implies that even very small asymmetric Rydberg interaction coefficients would be enough for such detection. We didn't estimate the interaction coefficient for asymmetric interaction here as that is out of the scope of the current work.

We didn't consider the effect of the decay of the Rydberg state on our alternative detection scheme yet. When the Rydberg state decays, the blockade effect goes away and there is no more suppression of photon emission. Hence, it seems best to detect early so that no Rydberg state decays. However, if we don't wait for long enough then there is not enough photons and we can't distinguish as seen above. Hence, there is an optimal time to stop detection which we found out to be 0.5 $\times$ decay time. Decreasing the detection time by half adversely affects the above estimate but not by a lot, especially given the small difference of photon number required in our scheme .

\subsection{Other Implementation Details}\label{sub_sec_other}

We now discuss several details of our proposal. We chose $nS$ Rydberg states because it offered two important simplifications in tracking the quantum dynamics. $nS$ states have only one angular momentum state $|n00\rangle$ and hence the product state $|n00\rangle|n00\rangle$ is the eigenstate of the non-interacting Hamiltonian \cite{walther2018interactions}. Also, owing to the zero angular momentum $nS$ states doesn't have any angle dependence in Rydberg interaction coefficients. Hence, for a polygon configuration of sites (Fig.~\ref{fig1}(C)) or even an arbitrary configuration (as explored in Fig.~\ref{fig3} for MIS problem) the Rydberg interaction and the corresponding evolution is tractable for $nS$ states (for upto 12 sites using computer systems available to us).

Although $nP$ states can be excited directly by one yellow laser and offer an experimentally simpler scenario, $nP$ states introduce considerable complications in tracking the quantum simulation. This includes angle-dependent Rydberg-Rydberg interaction and difficulty in exciting eigenstates of the Rydberg interaction Hamiltonian, which in this case are mostly entangled states \citep{walther2018interactions}.

For these reasons, we choose to address $nS$ states and take advantage of the additional flexibility in the geometry to apply this system to applications such as solving the MIS problem. However, the superposition states holds promise for potential non-trivial many-body quantum dynamics to be explored. In this context, many-body quantum dynamics of Rydberg exciton has recently been explored in \cite{poddubny2019topological}.

In the proposed experiment, we intend to excite the $ n$ = 25 excitation with about $1~\mu m$ radius and about $3~\mu m$ blockade radius. Hence to prevent double excitons occurring along the crystal propagation direction the crystal needs to be rather thin, around $3~\mu m$.  It is a challenge to manufacture such small thickness artificial Cu$_2$O crystals, although artificial Cu$_2$O micro-crystals has recently been created \cite{steinhauer2020rydberg}. Another problem with using a small thickness crystal is that it can change the energy level structure of the excitons, because it is now confined inside a potential well. Considering these issues, a little more thickness would actually probably be fine as the laser itself is focused in about $1-2~\mu m$ length at the centre of the crystal and hence weaker in intensity at the edges, resulting in larger blockade radius there. 

There is also a possibility that the excitons could move due to thermal motion despite the very small timescales used. This would be detrimental to our proposal. However, we estimated that in the simulation duration of $\sim$ 0.2 ns and at 1K temperature, even if an exciton moves ideally, uninhibited with its average thermal velocity, it would only travel $\sim 1 ~\mu m$. Moreover, in practical systems there is scattering due to crystal imperfection which would further slow down the exciton greatly. Also, the system can be cooled to mK temperatures to constrain the exciton montion even more.

\section{Conclusions}

In conclusion, we simulated many-body Rydberg excitation dynamics in Cu$_2$O excitons. Despite the detrimental effect of neighboring Rydberg states, we demonstrated that $\mathbb{Z}_2$-ordered phase can be reached and observed using correlation in detected emission. We explored the scaling of the success probability and showed that such many-body quantum states can be reached and observed for over 50 exciton sites at experimental run times two orders of magnitude shorter than the individually trapped Rydberg atoms. Using the ability to selectively excite excitons in an arbitrary configuration, we argued MIS problems can be solved in the Rydberg exciton system. To further scale up the system, we discussed the possibility of improving exciton lifetimes by many orders of magnitude using engineered 2D micro structures.  Overall these results show the potential of Cu$_2$O excitons and other attractive semiconductor systems for Rydberg excitons to be used to simulate many-body dynamics with applications extending to complex optimization problems.


\begin{thebibliography}{30}%
\makeatletter
\providecommand \@ifxundefined [1]{%
 \@ifx{#1\undefined}
}%
\providecommand \@ifnum [1]{%
 \ifnum #1\expandafter \@firstoftwo
 \else \expandafter \@secondoftwo
 \fi
}%
\providecommand \@ifx [1]{%
 \ifx #1\expandafter \@firstoftwo
 \else \expandafter \@secondoftwo
 \fi
}%
\providecommand \natexlab [1]{#1}%
\providecommand \enquote  [1]{``#1''}%
\providecommand \bibnamefont  [1]{#1}%
\providecommand \bibfnamefont [1]{#1}%
\providecommand \citenamefont [1]{#1}%
\providecommand \href@noop [0]{\@secondoftwo}%
\providecommand \href [0]{\begingroup \@sanitize@url \@href}%
\providecommand \@href[1]{\@@startlink{#1}\@@href}%
\providecommand \@@href[1]{\endgroup#1\@@endlink}%
\providecommand \@sanitize@url [0]{\catcode `\\12\catcode `\$12\catcode
  `\&12\catcode `\#12\catcode `\^12\catcode `\_12\catcode `\%12\relax}%
\providecommand \@@startlink[1]{}%
\providecommand \@@endlink[0]{}%
\providecommand \url  [0]{\begingroup\@sanitize@url \@url }%
\providecommand \@url [1]{\endgroup\@href {#1}{\urlprefix }}%
\providecommand \urlprefix  [0]{URL }%
\providecommand \Eprint [0]{\href }%
\providecommand \doibase [0]{http://dx.doi.org/}%
\providecommand \selectlanguage [0]{\@gobble}%
\providecommand \bibinfo  [0]{\@secondoftwo}%
\providecommand \bibfield  [0]{\@secondoftwo}%
\providecommand \translation [1]{[#1]}%
\providecommand \BibitemOpen [0]{}%
\providecommand \bibitemStop [0]{}%
\providecommand \bibitemNoStop [0]{.\EOS\space}%
\providecommand \EOS [0]{\spacefactor3000\relax}%
\providecommand \BibitemShut  [1]{\csname bibitem#1\endcsname}%
\let\auto@bib@innerbib\@empty
\bibitem [{\citenamefont {Saffman}\ \emph {et~al.}(2010)\citenamefont
  {Saffman}, \citenamefont {Walker},\ and\ \citenamefont
  {M{\o}lmer}}]{SaffmanRev}%
  \BibitemOpen
  \bibfield  {author} {\bibinfo {author} {\bibfnamefont {M.}~\bibnamefont
  {Saffman}}, \bibinfo {author} {\bibfnamefont {T.~G.}\ \bibnamefont {Walker}},
  \ and\ \bibinfo {author} {\bibfnamefont {K.}~\bibnamefont {M{\o}lmer}},\
  }\href@noop {} {\bibfield  {journal} {\bibinfo  {journal} {Reviews of Modern
  Physics}\ }\textbf {\bibinfo {volume} {82}},\ \bibinfo {pages} {2313}
  (\bibinfo {year} {2010})}\BibitemShut {NoStop}%
\bibitem [{\citenamefont {Šibalić}\ and\ \citenamefont
  {Adams}(2018)}]{SibalicAdamsBook}%
  \BibitemOpen
  \bibfield  {author} {\bibinfo {author} {\bibfnamefont {N.}~\bibnamefont
  {Šibalić}}\ and\ \bibinfo {author} {\bibfnamefont {C.~S.}\ \bibnamefont
  {Adams}},\ }\href@noop {} {\emph {\bibinfo {title} {Rydberg Physics}}},\
  2399-2891\ (\bibinfo  {publisher} {IOP Publishing},\ \bibinfo {year}
  {2018})\BibitemShut {NoStop}%
\bibitem [{\citenamefont {Urban}\ \emph {et~al.}(2009)\citenamefont {Urban},
  \citenamefont {Johnson}, \citenamefont {Henage}, \citenamefont {Isenhower},
  \citenamefont {Yavuz}, \citenamefont {Walker},\ and\ \citenamefont
  {Saffman}}]{Urban2009}%
  \BibitemOpen
  \bibfield  {author} {\bibinfo {author} {\bibfnamefont {E.}~\bibnamefont
  {Urban}}, \bibinfo {author} {\bibfnamefont {T.~A.}\ \bibnamefont {Johnson}},
  \bibinfo {author} {\bibfnamefont {T.}~\bibnamefont {Henage}}, \bibinfo
  {author} {\bibfnamefont {L.}~\bibnamefont {Isenhower}}, \bibinfo {author}
  {\bibfnamefont {D.}~\bibnamefont {Yavuz}}, \bibinfo {author} {\bibfnamefont
  {T.}~\bibnamefont {Walker}}, \ and\ \bibinfo {author} {\bibfnamefont
  {M.}~\bibnamefont {Saffman}},\ }\href@noop {} {\bibfield  {journal} {\bibinfo
   {journal} {Nature Physics}\ }\textbf {\bibinfo {volume} {5}},\ \bibinfo
  {pages} {110} (\bibinfo {year} {2009})}\BibitemShut {NoStop}%
\bibitem [{\citenamefont {Ga{\"e}tan}\ \emph {et~al.}(2009)\citenamefont
  {Ga{\"e}tan}, \citenamefont {Miroshnychenko}, \citenamefont {Wilk},
  \citenamefont {Chotia}, \citenamefont {Viteau}, \citenamefont {Comparat},
  \citenamefont {Pillet}, \citenamefont {Browaeys},\ and\ \citenamefont
  {Grangier}}]{Gaetan2009}%
  \BibitemOpen
  \bibfield  {author} {\bibinfo {author} {\bibfnamefont {A.}~\bibnamefont
  {Ga{\"e}tan}}, \bibinfo {author} {\bibfnamefont {Y.}~\bibnamefont
  {Miroshnychenko}}, \bibinfo {author} {\bibfnamefont {T.}~\bibnamefont
  {Wilk}}, \bibinfo {author} {\bibfnamefont {A.}~\bibnamefont {Chotia}},
  \bibinfo {author} {\bibfnamefont {M.}~\bibnamefont {Viteau}}, \bibinfo
  {author} {\bibfnamefont {D.}~\bibnamefont {Comparat}}, \bibinfo {author}
  {\bibfnamefont {P.}~\bibnamefont {Pillet}}, \bibinfo {author} {\bibfnamefont
  {A.}~\bibnamefont {Browaeys}}, \ and\ \bibinfo {author} {\bibfnamefont
  {P.}~\bibnamefont {Grangier}},\ }\href@noop {} {\bibfield  {journal}
  {\bibinfo  {journal} {Nature Physics}\ }\textbf {\bibinfo {volume} {5}},\
  \bibinfo {pages} {115} (\bibinfo {year} {2009})}\BibitemShut {NoStop}%
\bibitem [{\citenamefont {Peyronel}\ \emph {et~al.}(2012)\citenamefont
  {Peyronel}, \citenamefont {Firstenberg}, \citenamefont {Liang}, \citenamefont
  {Hofferberth}, \citenamefont {Gorshkov}, \citenamefont {Pohl}, \citenamefont
  {Lukin},\ and\ \citenamefont {Vuleti{\'c}}}]{peyronel2012quantum}%
  \BibitemOpen
  \bibfield  {author} {\bibinfo {author} {\bibfnamefont {T.}~\bibnamefont
  {Peyronel}}, \bibinfo {author} {\bibfnamefont {O.}~\bibnamefont
  {Firstenberg}}, \bibinfo {author} {\bibfnamefont {Q.-Y.}\ \bibnamefont
  {Liang}}, \bibinfo {author} {\bibfnamefont {S.}~\bibnamefont {Hofferberth}},
  \bibinfo {author} {\bibfnamefont {A.~V.}\ \bibnamefont {Gorshkov}}, \bibinfo
  {author} {\bibfnamefont {T.}~\bibnamefont {Pohl}}, \bibinfo {author}
  {\bibfnamefont {M.~D.}\ \bibnamefont {Lukin}}, \ and\ \bibinfo {author}
  {\bibfnamefont {V.}~\bibnamefont {Vuleti{\'c}}},\ }\href@noop {} {\bibfield
  {journal} {\bibinfo  {journal} {Nature}\ }\textbf {\bibinfo {volume} {488}},\
  \bibinfo {pages} {57} (\bibinfo {year} {2012})}\BibitemShut {NoStop}%
\bibitem [{\citenamefont {M{\"u}ller}\ \emph {et~al.}(2013)\citenamefont
  {M{\"u}ller}, \citenamefont {K{\"o}lle}, \citenamefont {L{\"o}w},
  \citenamefont {Pfau}, \citenamefont {Calarco},\ and\ \citenamefont
  {Montangero}}]{muller2013room}%
  \BibitemOpen
  \bibfield  {author} {\bibinfo {author} {\bibfnamefont {M.}~\bibnamefont
  {M{\"u}ller}}, \bibinfo {author} {\bibfnamefont {A.}~\bibnamefont
  {K{\"o}lle}}, \bibinfo {author} {\bibfnamefont {R.}~\bibnamefont {L{\"o}w}},
  \bibinfo {author} {\bibfnamefont {T.}~\bibnamefont {Pfau}}, \bibinfo {author}
  {\bibfnamefont {T.}~\bibnamefont {Calarco}}, \ and\ \bibinfo {author}
  {\bibfnamefont {S.}~\bibnamefont {Montangero}},\ }\href@noop {} {\bibfield
  {journal} {\bibinfo  {journal} {Physical Review A}\ }\textbf {\bibinfo
  {volume} {87}},\ \bibinfo {pages} {053412} (\bibinfo {year}
  {2013})}\BibitemShut {NoStop}%
\bibitem [{\citenamefont {Paredes-Barato}\ and\ \citenamefont
  {Adams}(2014)}]{paredes2014all}%
  \BibitemOpen
  \bibfield  {author} {\bibinfo {author} {\bibfnamefont {D.}~\bibnamefont
  {Paredes-Barato}}\ and\ \bibinfo {author} {\bibfnamefont {C.}~\bibnamefont
  {Adams}},\ }\href@noop {} {\bibfield  {journal} {\bibinfo  {journal}
  {Physical review letters}\ }\textbf {\bibinfo {volume} {112}},\ \bibinfo
  {pages} {040501} (\bibinfo {year} {2014})}\BibitemShut {NoStop}%
\bibitem [{\citenamefont {Gorniaczyk}\ \emph {et~al.}(2014)\citenamefont
  {Gorniaczyk}, \citenamefont {Tresp}, \citenamefont {Schmidt}, \citenamefont
  {Fedder},\ and\ \citenamefont {Hofferberth}}]{gorniaczyk2014single}%
  \BibitemOpen
  \bibfield  {author} {\bibinfo {author} {\bibfnamefont {H.}~\bibnamefont
  {Gorniaczyk}}, \bibinfo {author} {\bibfnamefont {C.}~\bibnamefont {Tresp}},
  \bibinfo {author} {\bibfnamefont {J.}~\bibnamefont {Schmidt}}, \bibinfo
  {author} {\bibfnamefont {H.}~\bibnamefont {Fedder}}, \ and\ \bibinfo {author}
  {\bibfnamefont {S.}~\bibnamefont {Hofferberth}},\ }\href@noop {} {\bibfield
  {journal} {\bibinfo  {journal} {Physical review letters}\ }\textbf {\bibinfo
  {volume} {113}},\ \bibinfo {pages} {053601} (\bibinfo {year}
  {2014})}\BibitemShut {NoStop}%
\bibitem [{\citenamefont {K{\'o}m{\'a}r}\ \emph {et~al.}(2016)\citenamefont
  {K{\'o}m{\'a}r}, \citenamefont {Topcu}, \citenamefont {Kessler},
  \citenamefont {Derevianko}, \citenamefont {Vuleti{\'c}}, \citenamefont {Ye},\
  and\ \citenamefont {Lukin}}]{komar2016quantum}%
  \BibitemOpen
  \bibfield  {author} {\bibinfo {author} {\bibfnamefont {P.}~\bibnamefont
  {K{\'o}m{\'a}r}}, \bibinfo {author} {\bibfnamefont {T.}~\bibnamefont
  {Topcu}}, \bibinfo {author} {\bibfnamefont {E.}~\bibnamefont {Kessler}},
  \bibinfo {author} {\bibfnamefont {A.}~\bibnamefont {Derevianko}}, \bibinfo
  {author} {\bibfnamefont {V.}~\bibnamefont {Vuleti{\'c}}}, \bibinfo {author}
  {\bibfnamefont {J.}~\bibnamefont {Ye}}, \ and\ \bibinfo {author}
  {\bibfnamefont {M.}~\bibnamefont {Lukin}},\ }\href@noop {} {\bibfield
  {journal} {\bibinfo  {journal} {Physical review letters}\ }\textbf {\bibinfo
  {volume} {117}},\ \bibinfo {pages} {060506} (\bibinfo {year}
  {2016})}\BibitemShut {NoStop}%
\bibitem [{\citenamefont {Han}\ \emph {et~al.}(2010)\citenamefont {Han},
  \citenamefont {He}, \citenamefont {Heshami}, \citenamefont {Li},\ and\
  \citenamefont {Simon}}]{han2010quantum}%
  \BibitemOpen
  \bibfield  {author} {\bibinfo {author} {\bibfnamefont {Y.}~\bibnamefont
  {Han}}, \bibinfo {author} {\bibfnamefont {B.}~\bibnamefont {He}}, \bibinfo
  {author} {\bibfnamefont {K.}~\bibnamefont {Heshami}}, \bibinfo {author}
  {\bibfnamefont {C.-Z.}\ \bibnamefont {Li}}, \ and\ \bibinfo {author}
  {\bibfnamefont {C.}~\bibnamefont {Simon}},\ }\href@noop {} {\bibfield
  {journal} {\bibinfo  {journal} {Physical Review A}\ }\textbf {\bibinfo
  {volume} {81}},\ \bibinfo {pages} {052311} (\bibinfo {year}
  {2010})}\BibitemShut {NoStop}%
\bibitem [{\citenamefont {Zhao}\ \emph {et~al.}(2010)\citenamefont {Zhao},
  \citenamefont {M\"uller}, \citenamefont {Hammerer},\ and\ \citenamefont
  {Zoller}}]{PhysRevA.81.052329}%
  \BibitemOpen
  \bibfield  {author} {\bibinfo {author} {\bibfnamefont {B.}~\bibnamefont
  {Zhao}}, \bibinfo {author} {\bibfnamefont {M.}~\bibnamefont {M\"uller}},
  \bibinfo {author} {\bibfnamefont {K.}~\bibnamefont {Hammerer}}, \ and\
  \bibinfo {author} {\bibfnamefont {P.}~\bibnamefont {Zoller}},\ }\href
  {\doibase 10.1103/PhysRevA.81.052329} {\bibfield  {journal} {\bibinfo
  {journal} {Phys. Rev. A}\ }\textbf {\bibinfo {volume} {81}},\ \bibinfo
  {pages} {052329} (\bibinfo {year} {2010})}\BibitemShut {NoStop}%
\bibitem [{\citenamefont {Bernien}\ \emph {et~al.}(2017)\citenamefont
  {Bernien}, \citenamefont {Schwartz}, \citenamefont {Keesling}, \citenamefont
  {Levine}, \citenamefont {Omran}, \citenamefont {Pichler}, \citenamefont
  {Choi}, \citenamefont {Zibrov}, \citenamefont {Endres}, \citenamefont
  {Greiner} \emph {et~al.}}]{bernien2017probing}%
  \BibitemOpen
  \bibfield  {author} {\bibinfo {author} {\bibfnamefont {H.}~\bibnamefont
  {Bernien}}, \bibinfo {author} {\bibfnamefont {S.}~\bibnamefont {Schwartz}},
  \bibinfo {author} {\bibfnamefont {A.}~\bibnamefont {Keesling}}, \bibinfo
  {author} {\bibfnamefont {H.}~\bibnamefont {Levine}}, \bibinfo {author}
  {\bibfnamefont {A.}~\bibnamefont {Omran}}, \bibinfo {author} {\bibfnamefont
  {H.}~\bibnamefont {Pichler}}, \bibinfo {author} {\bibfnamefont
  {S.}~\bibnamefont {Choi}}, \bibinfo {author} {\bibfnamefont {A.~S.}\
  \bibnamefont {Zibrov}}, \bibinfo {author} {\bibfnamefont {M.}~\bibnamefont
  {Endres}}, \bibinfo {author} {\bibfnamefont {M.}~\bibnamefont {Greiner}},
  \emph {et~al.},\ }\href@noop {} {\bibfield  {journal} {\bibinfo  {journal}
  {Nature}\ }\textbf {\bibinfo {volume} {551}},\ \bibinfo {pages} {579}
  (\bibinfo {year} {2017})}\BibitemShut {NoStop}%
\bibitem [{\citenamefont {Keesling}\ \emph {et~al.}(2019)\citenamefont
  {Keesling}, \citenamefont {Omran}, \citenamefont {Levine}, \citenamefont
  {Bernien}, \citenamefont {Pichler}, \citenamefont {Choi}, \citenamefont
  {Samajdar}, \citenamefont {Schwartz}, \citenamefont {Silvi}, \citenamefont
  {Sachdev} \emph {et~al.}}]{keesling2019quantum}%
  \BibitemOpen
  \bibfield  {author} {\bibinfo {author} {\bibfnamefont {A.}~\bibnamefont
  {Keesling}}, \bibinfo {author} {\bibfnamefont {A.}~\bibnamefont {Omran}},
  \bibinfo {author} {\bibfnamefont {H.}~\bibnamefont {Levine}}, \bibinfo
  {author} {\bibfnamefont {H.}~\bibnamefont {Bernien}}, \bibinfo {author}
  {\bibfnamefont {H.}~\bibnamefont {Pichler}}, \bibinfo {author} {\bibfnamefont
  {S.}~\bibnamefont {Choi}}, \bibinfo {author} {\bibfnamefont {R.}~\bibnamefont
  {Samajdar}}, \bibinfo {author} {\bibfnamefont {S.}~\bibnamefont {Schwartz}},
  \bibinfo {author} {\bibfnamefont {P.}~\bibnamefont {Silvi}}, \bibinfo
  {author} {\bibfnamefont {S.}~\bibnamefont {Sachdev}},  \emph {et~al.},\
  }\href@noop {} {\bibfield  {journal} {\bibinfo  {journal} {Nature}\ }\textbf
  {\bibinfo {volume} {568}},\ \bibinfo {pages} {207} (\bibinfo {year}
  {2019})}\BibitemShut {NoStop}%
\bibitem [{\citenamefont {Omran}\ \emph {et~al.}(2019)\citenamefont {Omran},
  \citenamefont {Levine}, \citenamefont {Keesling}, \citenamefont {Semeghini},
  \citenamefont {Wang}, \citenamefont {Ebadi}, \citenamefont {Bernien},
  \citenamefont {Zibrov}, \citenamefont {Pichler}, \citenamefont {Choi} \emph
  {et~al.}}]{omran2019generation}%
  \BibitemOpen
  \bibfield  {author} {\bibinfo {author} {\bibfnamefont {A.}~\bibnamefont
  {Omran}}, \bibinfo {author} {\bibfnamefont {H.}~\bibnamefont {Levine}},
  \bibinfo {author} {\bibfnamefont {A.}~\bibnamefont {Keesling}}, \bibinfo
  {author} {\bibfnamefont {G.}~\bibnamefont {Semeghini}}, \bibinfo {author}
  {\bibfnamefont {T.~T.}\ \bibnamefont {Wang}}, \bibinfo {author}
  {\bibfnamefont {S.}~\bibnamefont {Ebadi}}, \bibinfo {author} {\bibfnamefont
  {H.}~\bibnamefont {Bernien}}, \bibinfo {author} {\bibfnamefont {A.~S.}\
  \bibnamefont {Zibrov}}, \bibinfo {author} {\bibfnamefont {H.}~\bibnamefont
  {Pichler}}, \bibinfo {author} {\bibfnamefont {S.}~\bibnamefont {Choi}},
  \emph {et~al.},\ }\href@noop {} {\bibfield  {journal} {\bibinfo  {journal}
  {Science}\ }\textbf {\bibinfo {volume} {365}},\ \bibinfo {pages} {570}
  (\bibinfo {year} {2019})}\BibitemShut {NoStop}%
\bibitem [{\citenamefont {Pichler}\ \emph
  {et~al.}(2018{\natexlab{a}})\citenamefont {Pichler}, \citenamefont {Wang},
  \citenamefont {Zhou}, \citenamefont {Choi},\ and\ \citenamefont
  {Lukin}}]{pichler2018quantum}%
  \BibitemOpen
  \bibfield  {author} {\bibinfo {author} {\bibfnamefont {H.}~\bibnamefont
  {Pichler}}, \bibinfo {author} {\bibfnamefont {S.-T.}\ \bibnamefont {Wang}},
  \bibinfo {author} {\bibfnamefont {L.}~\bibnamefont {Zhou}}, \bibinfo {author}
  {\bibfnamefont {S.}~\bibnamefont {Choi}}, \ and\ \bibinfo {author}
  {\bibfnamefont {M.~D.}\ \bibnamefont {Lukin}},\ }\href@noop {} {\bibfield
  {journal} {\bibinfo  {journal} {arXiv preprint arXiv:1808.10816}\ } (\bibinfo
  {year} {2018}{\natexlab{a}})}\BibitemShut {NoStop}%
\bibitem [{\citenamefont {Kazimierczuk}\ \emph {et~al.}(2014)\citenamefont
  {Kazimierczuk}, \citenamefont {Fr{\"o}hlich}, \citenamefont {Scheel},
  \citenamefont {Stolz},\ and\ \citenamefont {Bayer}}]{kazimierczuk2014giant}%
  \BibitemOpen
  \bibfield  {author} {\bibinfo {author} {\bibfnamefont {T.}~\bibnamefont
  {Kazimierczuk}}, \bibinfo {author} {\bibfnamefont {D.}~\bibnamefont
  {Fr{\"o}hlich}}, \bibinfo {author} {\bibfnamefont {S.}~\bibnamefont
  {Scheel}}, \bibinfo {author} {\bibfnamefont {H.}~\bibnamefont {Stolz}}, \
  and\ \bibinfo {author} {\bibfnamefont {M.}~\bibnamefont {Bayer}},\
  }\href@noop {} {\bibfield  {journal} {\bibinfo  {journal} {Nature}\ }\textbf
  {\bibinfo {volume} {514}},\ \bibinfo {pages} {343} (\bibinfo {year}
  {2014})}\BibitemShut {NoStop}%
\bibitem [{\citenamefont {Khazali}\ \emph {et~al.}(2017)\citenamefont
  {Khazali}, \citenamefont {Heshami},\ and\ \citenamefont
  {Simon}}]{khazali2017single}%
  \BibitemOpen
  \bibfield  {author} {\bibinfo {author} {\bibfnamefont {M.}~\bibnamefont
  {Khazali}}, \bibinfo {author} {\bibfnamefont {K.}~\bibnamefont {Heshami}}, \
  and\ \bibinfo {author} {\bibfnamefont {C.}~\bibnamefont {Simon}},\
  }\href@noop {} {\bibfield  {journal} {\bibinfo  {journal} {Journal of Physics
  B: Atomic, Molecular and Optical Physics}\ }\textbf {\bibinfo {volume}
  {50}},\ \bibinfo {pages} {215301} (\bibinfo {year} {2017})}\BibitemShut
  {NoStop}%
\bibitem [{\citenamefont {Walther}\ \emph
  {et~al.}(2018{\natexlab{a}})\citenamefont {Walther}, \citenamefont {Johne},\
  and\ \citenamefont {Pohl}}]{walther2018giant}%
  \BibitemOpen
  \bibfield  {author} {\bibinfo {author} {\bibfnamefont {V.}~\bibnamefont
  {Walther}}, \bibinfo {author} {\bibfnamefont {R.}~\bibnamefont {Johne}}, \
  and\ \bibinfo {author} {\bibfnamefont {T.}~\bibnamefont {Pohl}},\ }\href@noop
  {} {\bibfield  {journal} {\bibinfo  {journal} {Nature communications}\
  }\textbf {\bibinfo {volume} {9}},\ \bibinfo {pages} {1} (\bibinfo {year}
  {2018}{\natexlab{a}})}\BibitemShut {NoStop}%
\bibitem [{\citenamefont {Poddubny}\ and\ \citenamefont
  {Glazov}(2019)}]{poddubny2019topological}%
  \BibitemOpen
  \bibfield  {author} {\bibinfo {author} {\bibfnamefont {A.}~\bibnamefont
  {Poddubny}}\ and\ \bibinfo {author} {\bibfnamefont {M.}~\bibnamefont
  {Glazov}},\ }\href@noop {} {\bibfield  {journal} {\bibinfo  {journal}
  {Physical review letters}\ }\textbf {\bibinfo {volume} {123}},\ \bibinfo
  {pages} {126801} (\bibinfo {year} {2019})}\BibitemShut {NoStop}%
\bibitem [{\citenamefont {Uihlein}\ \emph
  {et~al.}(1981{\natexlab{a}})\citenamefont {Uihlein}, \citenamefont
  {Fr\"ohlich},\ and\ \citenamefont {Kenklies}}]{PhysRevB.23.2731}%
  \BibitemOpen
  \bibfield  {author} {\bibinfo {author} {\bibfnamefont {C.}~\bibnamefont
  {Uihlein}}, \bibinfo {author} {\bibfnamefont {D.}~\bibnamefont {Fr\"ohlich}},
  \ and\ \bibinfo {author} {\bibfnamefont {R.}~\bibnamefont {Kenklies}},\
  }\href {\doibase 10.1103/PhysRevB.23.2731} {\bibfield  {journal} {\bibinfo
  {journal} {Phys. Rev. B}\ }\textbf {\bibinfo {volume} {23}},\ \bibinfo
  {pages} {2731} (\bibinfo {year} {1981}{\natexlab{a}})}\BibitemShut {NoStop}%
\bibitem [{\citenamefont {Walther}\ \emph {et~al.}(2020)\citenamefont
  {Walther}, \citenamefont {Gr\"unwald},\ and\ \citenamefont
  {Pohl}}]{PhysRevLett.125.173601}%
  \BibitemOpen
  \bibfield  {author} {\bibinfo {author} {\bibfnamefont {V.}~\bibnamefont
  {Walther}}, \bibinfo {author} {\bibfnamefont {P.}~\bibnamefont {Gr\"unwald}},
  \ and\ \bibinfo {author} {\bibfnamefont {T.}~\bibnamefont {Pohl}},\ }\href
  {\doibase 10.1103/PhysRevLett.125.173601} {\bibfield  {journal} {\bibinfo
  {journal} {Phys. Rev. Lett.}\ }\textbf {\bibinfo {volume} {125}},\ \bibinfo
  {pages} {173601} (\bibinfo {year} {2020})}\BibitemShut {NoStop}%
\bibitem [{\citenamefont {Sch\"one}\ \emph {et~al.}(2017)\citenamefont
  {Sch\"one}, \citenamefont {Stolz},\ and\ \citenamefont
  {Naka}}]{PhysRevB.96.115207}%
  \BibitemOpen
  \bibfield  {author} {\bibinfo {author} {\bibfnamefont {F.}~\bibnamefont
  {Sch\"one}}, \bibinfo {author} {\bibfnamefont {H.}~\bibnamefont {Stolz}}, \
  and\ \bibinfo {author} {\bibfnamefont {N.}~\bibnamefont {Naka}},\ }\href
  {\doibase 10.1103/PhysRevB.96.115207} {\bibfield  {journal} {\bibinfo
  {journal} {Phys. Rev. B}\ }\textbf {\bibinfo {volume} {96}},\ \bibinfo
  {pages} {115207} (\bibinfo {year} {2017})}\BibitemShut {NoStop}%
\bibitem [{\citenamefont {Uihlein}\ \emph
  {et~al.}(1981{\natexlab{b}})\citenamefont {Uihlein}, \citenamefont
  {Fr{\"o}hlich},\ and\ \citenamefont {Kenklies}}]{uihlein1981investigation}%
  \BibitemOpen
  \bibfield  {author} {\bibinfo {author} {\bibfnamefont {C.}~\bibnamefont
  {Uihlein}}, \bibinfo {author} {\bibfnamefont {D.}~\bibnamefont
  {Fr{\"o}hlich}}, \ and\ \bibinfo {author} {\bibfnamefont {R.}~\bibnamefont
  {Kenklies}},\ }\href@noop {} {\bibfield  {journal} {\bibinfo  {journal}
  {Physical Review B}\ }\textbf {\bibinfo {volume} {23}},\ \bibinfo {pages}
  {2731} (\bibinfo {year} {1981}{\natexlab{b}})}\BibitemShut {NoStop}%
\bibitem [{\citenamefont {Heck{\"o}tter}\ \emph {et~al.}(2020)\citenamefont
  {Heck{\"o}tter}, \citenamefont {Walther}, \citenamefont {Scheel},
  \citenamefont {Bayer}, \citenamefont {Pohl},\ and\ \citenamefont
  {A{\ss}mann}}]{heckotter2020asymmetric}%
  \BibitemOpen
  \bibfield  {author} {\bibinfo {author} {\bibfnamefont {J.}~\bibnamefont
  {Heck{\"o}tter}}, \bibinfo {author} {\bibfnamefont {V.}~\bibnamefont
  {Walther}}, \bibinfo {author} {\bibfnamefont {S.}~\bibnamefont {Scheel}},
  \bibinfo {author} {\bibfnamefont {M.}~\bibnamefont {Bayer}}, \bibinfo
  {author} {\bibfnamefont {T.}~\bibnamefont {Pohl}}, \ and\ \bibinfo {author}
  {\bibfnamefont {M.}~\bibnamefont {A{\ss}mann}},\ }\href@noop {} {\bibfield
  {journal} {\bibinfo  {journal} {arXiv preprint arXiv:2010.15459}\ } (\bibinfo
  {year} {2020})}\BibitemShut {NoStop}%
\bibitem [{\citenamefont {Walther}\ \emph
  {et~al.}(2018{\natexlab{b}})\citenamefont {Walther}, \citenamefont
  {Kr{\"u}ger}, \citenamefont {Scheel},\ and\ \citenamefont
  {Pohl}}]{walther2018interactions}%
  \BibitemOpen
  \bibfield  {author} {\bibinfo {author} {\bibfnamefont {V.}~\bibnamefont
  {Walther}}, \bibinfo {author} {\bibfnamefont {S.~O.}\ \bibnamefont
  {Kr{\"u}ger}}, \bibinfo {author} {\bibfnamefont {S.}~\bibnamefont {Scheel}},
  \ and\ \bibinfo {author} {\bibfnamefont {T.}~\bibnamefont {Pohl}},\
  }\href@noop {} {\bibfield  {journal} {\bibinfo  {journal} {Physical Review
  B}\ }\textbf {\bibinfo {volume} {98}},\ \bibinfo {pages} {165201} (\bibinfo
  {year} {2018}{\natexlab{b}})}\BibitemShut {NoStop}%
\bibitem [{\citenamefont {Pichler}\ \emph
  {et~al.}(2018{\natexlab{b}})\citenamefont {Pichler}, \citenamefont {Wang},
  \citenamefont {Zhou}, \citenamefont {Choi},\ and\ \citenamefont
  {Lukin}}]{pichler2018computational}%
  \BibitemOpen
  \bibfield  {author} {\bibinfo {author} {\bibfnamefont {H.}~\bibnamefont
  {Pichler}}, \bibinfo {author} {\bibfnamefont {S.-T.}\ \bibnamefont {Wang}},
  \bibinfo {author} {\bibfnamefont {L.}~\bibnamefont {Zhou}}, \bibinfo {author}
  {\bibfnamefont {S.}~\bibnamefont {Choi}}, \ and\ \bibinfo {author}
  {\bibfnamefont {M.~D.}\ \bibnamefont {Lukin}},\ }\href@noop {} {\bibfield
  {journal} {\bibinfo  {journal} {arXiv preprint arXiv:1809.04954}\ } (\bibinfo
  {year} {2018}{\natexlab{b}})}\BibitemShut {NoStop}%
\bibitem [{\citenamefont {Wang}\ \emph {et~al.}(2019)\citenamefont {Wang},
  \citenamefont {Rhodes}, \citenamefont {Watanabe}, \citenamefont {Taniguchi},
  \citenamefont {Hone}, \citenamefont {Shan},\ and\ \citenamefont
  {Mak}}]{wang2019evidence}%
  \BibitemOpen
  \bibfield  {author} {\bibinfo {author} {\bibfnamefont {Z.}~\bibnamefont
  {Wang}}, \bibinfo {author} {\bibfnamefont {D.~A.}\ \bibnamefont {Rhodes}},
  \bibinfo {author} {\bibfnamefont {K.}~\bibnamefont {Watanabe}}, \bibinfo
  {author} {\bibfnamefont {T.}~\bibnamefont {Taniguchi}}, \bibinfo {author}
  {\bibfnamefont {J.~C.}\ \bibnamefont {Hone}}, \bibinfo {author}
  {\bibfnamefont {J.}~\bibnamefont {Shan}}, \ and\ \bibinfo {author}
  {\bibfnamefont {K.~F.}\ \bibnamefont {Mak}},\ }\href@noop {} {\bibfield
  {journal} {\bibinfo  {journal} {Nature}\ }\textbf {\bibinfo {volume} {574}},\
  \bibinfo {pages} {76} (\bibinfo {year} {2019})}\BibitemShut {NoStop}%
\bibitem [{\citenamefont {Montblanch}\ \emph {et~al.}(2020)\citenamefont
  {Montblanch}, \citenamefont {Kara}, \citenamefont {Paradeisanos},
  \citenamefont {Purser}, \citenamefont {Wang}, \citenamefont {Latawiec},
  \citenamefont {Loncar}, \citenamefont {Tongay}, \citenamefont {Ferrari},\
  and\ \citenamefont {Atature}}]{montblanch2020confinement}%
  \BibitemOpen
  \bibfield  {author} {\bibinfo {author} {\bibfnamefont {A.}~\bibnamefont
  {Montblanch}}, \bibinfo {author} {\bibfnamefont {D.}~\bibnamefont {Kara}},
  \bibinfo {author} {\bibfnamefont {I.}~\bibnamefont {Paradeisanos}}, \bibinfo
  {author} {\bibfnamefont {C.}~\bibnamefont {Purser}}, \bibinfo {author}
  {\bibfnamefont {G.}~\bibnamefont {Wang}}, \bibinfo {author} {\bibfnamefont
  {P.}~\bibnamefont {Latawiec}}, \bibinfo {author} {\bibfnamefont
  {M.}~\bibnamefont {Loncar}}, \bibinfo {author} {\bibfnamefont
  {S.}~\bibnamefont {Tongay}}, \bibinfo {author} {\bibfnamefont
  {A.}~\bibnamefont {Ferrari}}, \ and\ \bibinfo {author} {\bibfnamefont
  {M.}~\bibnamefont {Atature}},\ }\href@noop {} {\bibfield  {journal} {\bibinfo
   {journal} {Bulletin of the American Physical Society}\ } (\bibinfo {year}
  {2020})}\BibitemShut {NoStop}%
\bibitem [{\citenamefont {Kr{\"u}ger}\ and\ \citenamefont
  {Scheel}(2018)}]{kruger2018waveguides}%
  \BibitemOpen
  \bibfield  {author} {\bibinfo {author} {\bibfnamefont {S.~O.}\ \bibnamefont
  {Kr{\"u}ger}}\ and\ \bibinfo {author} {\bibfnamefont {S.}~\bibnamefont
  {Scheel}},\ }\href@noop {} {\bibfield  {journal} {\bibinfo  {journal}
  {Physical Review B}\ }\textbf {\bibinfo {volume} {97}},\ \bibinfo {pages}
  {205208} (\bibinfo {year} {2018})}\BibitemShut {NoStop}%
\bibitem [{\citenamefont {Steinhauer}\ \emph {et~al.}(2020)\citenamefont
  {Steinhauer}, \citenamefont {Versteegh}, \citenamefont {Gyger}, \citenamefont
  {Elshaari}, \citenamefont {Kunert}, \citenamefont {Mysyrowicz},\ and\
  \citenamefont {Zwiller}}]{steinhauer2020rydberg}%
  \BibitemOpen
  \bibfield  {author} {\bibinfo {author} {\bibfnamefont {S.}~\bibnamefont
  {Steinhauer}}, \bibinfo {author} {\bibfnamefont {M.~A.}\ \bibnamefont
  {Versteegh}}, \bibinfo {author} {\bibfnamefont {S.}~\bibnamefont {Gyger}},
  \bibinfo {author} {\bibfnamefont {A.~W.}\ \bibnamefont {Elshaari}}, \bibinfo
  {author} {\bibfnamefont {B.}~\bibnamefont {Kunert}}, \bibinfo {author}
  {\bibfnamefont {A.}~\bibnamefont {Mysyrowicz}}, \ and\ \bibinfo {author}
  {\bibfnamefont {V.}~\bibnamefont {Zwiller}},\ }\href@noop {} {\bibfield
  {journal} {\bibinfo  {journal} {Communications Materials}\ }\textbf {\bibinfo
  {volume} {1}},\ \bibinfo {pages} {1} (\bibinfo {year} {2020})}\BibitemShut
  {NoStop}%
\end{thebibliography}
%

\end{document}